# Relativistic particle in the rest frame

V. N. Salomatov

Irkutsk State Transport University,

15 Tchernyshevsky st., 664074 Irkutsk, Russia

E-mail: sav@irgups.ru

**Abstract**

It is shown that for a specific choice of a particular solution of the relativistic wave equation, it falls into the Helmholtz equation and the Klein-Gordon equation. In this case, the squares of the rest masses of the particle with the relativistic dispersion relation are determined by the Helmholtz equation. It is suggested the possibility in principle of classification of the mass spectrum and lifetimes of the particles according to the discrete solutions of the Helmholtz equation in a particular model. In this case, it can be assumed that the lifetime of a particle is determined by the ratio of contributions to the wave function of the Bessel and collapsed Neumann functions. An opportunity to compare real elementary particles rest masses and especially lifetimes with the particles masses and lifetimes considered is discussed.

**Introduction**

Despite the great success of the standard model [1, 2] due to the Higgs boson observation, there are difficulties in the description of the exact values of the rest masses and especially the lifetimes of elementary particles. In this regard, it is of interest to consider the additional features of quantum field theory (QFT) [3-5] using a simple coordinate-time representation of wave functions (Dirac-Weyl representation). The dynamic properties of a particle as in QFT itself so e.g. in its applications to the quantum theory of solids [6] are determined by the dispersion relation $E = E(\mathbf{k})$ (E-the energy, $\mathbf{k}$- the wave vector), and the velocity of particle, interpreted as the group velocity of the wave packet is given by

$$\mathbf{v} = \frac{1}{\hbar} \frac{\partial E}{\partial \mathbf{k}}, \qquad (1)$$

where $\hbar$ is the Planck constant.

This study is an attempt to consider a particle, which wave function satisfies the relativistic wave equation with relativistic dispersion

$$E = \pm c\hbar \left(\frac{mc^2}{\hbar^2} + \kappa^2\right)^{1/2}, \qquad (2)$$

where c is a light speed in a vacuum, m - the rest mass of the particle.

In this case, attention is drawn to the following formal circumstance. Among the particular solutions of relativistic wave equation, there are solutions that can be represented as a product of two functions. One of these functions satisfies the Helmholtz equation, which determines the rest mass of the particle, and the other satisfies the Klein -Gordon equation.

### Klein-Gordon and Helmholtz equations as consequences of the wave equation

Let's write the relativistic wave equation in the form

$$\left[\hat{p}^2 + \frac{\hbar^2}{c^2}\frac{\partial^2}{\partial t^2}\right]\Psi = 0, \qquad (3)$$

where $\hat{p}^2 = (-i\hbar\nabla)^2$ is the momentum operator, $\nabla$ - nabla operator.

Solutions of this linear equation are linear combinations (or Fourier integrals) of functions

$$\Psi_k = N_k \tilde{u} \exp\left[i\left(\kappa r - \frac{E}{\hbar}\right)t\right], \qquad (4)$$

where $\tilde{u}$ is independent of the coordinates and time the spin function two-component for bosons and four-component for fermions. One particular type of such combinations may be represented as the product of two functions

$$\Psi_{q,k} = \frac{N_{q,k}}{N_k}\left\{\exp\left[i(\kappa+q)r - \frac{E}{\hbar}t\right] \pm \exp\left[i(\kappa-q)r - \frac{E}{\hbar}t\right]\right\} = \Psi_q \Psi_\kappa. \qquad (5)$$

Here $N_\kappa, N_{q,\kappa}, N_q = \frac{N_{q,\kappa}}{N_\kappa}$ - normalization factors,

$$\Psi_q = N_q\left(e^{iqr} \pm e^{iqr}\right), \qquad (6)$$

**q** – vector perpendicular **k**. Note that in the case **q**⊥**k** the relations

$$(\hat{p}\,\Psi_q)\hbar\kappa = 0, \quad (\hat{p}\,\Psi_q)(\hat{p}_k\Psi) = 0, \qquad (7)$$

$$\langle \Psi_{q,\kappa}|\hat{p}|\Psi_{q,\kappa}\rangle = \langle \Psi_k|\hat{p}|\Psi_\kappa\rangle = \hat{p}\,\Psi_k = \hbar\kappa \qquad (8)$$

are carried out.

Let's generalize the problem for the form of the $\Psi_q$ function with the equal rights of all the perpendicular vector **k** directions. We seek a particular solution of equation (3) as follows

$$\Psi_{q,\kappa} = \Psi_q \Psi_k, \qquad (9)$$

where $\Psi_\kappa$ has the form (4).

Substituting (9) in (3), we have

$$\Psi_\kappa\left[\hat{p}^2\,\Psi_q + 2(\hat{p}\,\Psi_q)\hbar\kappa + \hbar^2\kappa^2\Psi_q - \frac{E^2}{c^2}\Psi_q\right] = 0. \qquad (10)$$

The term in square brackets is equal to zero when the condition (7) is performed and the equation

$$\hat{p}^2\,\Psi_q = \hbar^2 q^2\,\Psi_q \qquad (11)$$

is executed and the dispersion relation

$$E = \pm c\hbar(q^2 + \kappa^2)^{1/2} \qquad (12)$$

is fulfilled.

On the other hand, substituting (9) into (3) and performing the division of both sides of the equation by $\Psi_q$, if (7) and (11) are carried out, we obtain the equation for the $\Psi_\kappa$ function

$$\left[\hat{p}^2 + \frac{\hbar^2}{c^2}\frac{\partial^2}{\partial t^2} + \hbar^2\mathbf{q}^2\right]\Psi_k = 0, \qquad (13)$$

having the form of the Klein - Gordon equation.

The function (4) satisfies this equation for the implementation of the dispersion (12). The dispersion relation (12) coincides with the relativistic dispersion relation (2) if to the $\mathbf{q}^2$ value to make the sense of the $\dfrac{m^2 c^2}{\hbar^2}$ value, that is, if you believe

$$\mathbf{q}^2 = \frac{m^2 c^2}{\hbar^2}. \qquad (14)$$

In this case, the square of the rest mass is an $\hat{p}^2/c^2$ operator eigenvalue.

## The polar model of a particle in the rest frame

Rest frame for the particle can be considered as a frame in which the velocity is zero. When the dispersion relation (2, 12) is carried out from (1) we have

$$\mathbf{v} = c\mathbf{k}\left(\mathbf{q}^2 + \boldsymbol{\kappa}^2\right)^{-1/2}, \qquad (15)$$

i.e. $\mathbf{v} \to 0$ in the frame where $\boldsymbol{\kappa} \to 0$.

In QFT, e.g. [3-6], wave function of a free particle is given by (4). When $\boldsymbol{\kappa} \to 0$ with consideration of the wave function (4) does not make sense. At the same time, the particle wave function (5, 9) when $\boldsymbol{\kappa} \to 0$ takes the form

$$N_\mathbf{k}\, \Psi_\mathbf{q}\, \tilde{u}\, \exp\left(-i\frac{E}{\hbar}t\right) \qquad (16)$$

and can be regarded as particle wave function in the rest frame, and the specific form of this function is determined by the condition (7) and equation (11).

Solutions of the equation (11) are determined by the symmetry of the problem, the appropriate choice of the coordinate system (e.g, Cartesian, spherical, cylindrical) and by the form of the boundary conditions [7-9]. In an arbitrary frame ($\mathbf{k} \neq 0$), the wave function (9) has cylindrical symmetry with a symmetry axis along a $\boldsymbol{\kappa}$ vector. This symmetry is not violated if the $\Psi_\mathbf{q}$ function in the rest frame has a spherical or cylindrical symmetry.

Let's consider the simplest case in the rest frame, when $\Psi_\mathbf{q}$ has a polar symmetry, i. e, we consider the solution of equation (11) in the polar coordinate system ($\rho, \varphi$). For the condition

(7) is enough to put that in an arbitrary reference frame ($\mathbf{k} \neq 0$) the ($\rho, \varphi$) plane is perpendicular to the $\mathbf{\kappa}$ vector. After separation of the radial and azimuthal variables

$$\Psi_{\mathbf{q}} = R(\rho) \cdot \Phi(\varphi) \qquad (17)$$

the equation for $R(\rho)$ becomes

$$\frac{\partial^2 R}{\partial \rho^2} + \frac{1}{\rho} \frac{\partial R}{\partial \rho} - \frac{n^2}{\rho^2} R + q^2 R = 0. \qquad (18)$$

After the transformation $q\rho = x$ equation (18) reduces to the n - th order Bessel equation

$$\frac{\partial^2 F}{\partial x^2} + \frac{1}{x} \frac{\partial F}{\partial x} + \left(1 - \frac{n^2}{x^2}\right) F = 0, \qquad (19)$$

where F(x)=R(x/q).

The general solution of (19) is a linear combination of Bessel and Neumann functions and modified Bessel and Neumann functions. When $q^2 > 0$ the general solution of (19) is a linear combination of the Bessel and Neumann functions [7-9]

$$R_n(r) = A N_n(r) + B J_n(r). \qquad (20)$$

Here, $J_n(x)$ is the Bessel function, $N_n(x)$ is the Neumann function, $A$ and $B$ are expansion coefficients determined from the boundary conditions and the normalization condition.

Implementation of the boundary conditions used, for example, in vibration theory and in the theory of heat conduction [7-9]

$$R(a) = 0, \qquad (21)$$

as well as the $R(\rho)$ continuity and limitation conditions under $\rho \to 0$ lead to the vanishing of the $A$ coefficients.

Equations (18.19), when (21) is carried out, have solutions for discrete values of $q^2$ magnitude, that is, when

$$q_{nl}^2 = \frac{1}{a^2} X_{nl}^2, \qquad (22)$$

wherein $X_{nl}$ are values of the roots of (20), n = 0, 1, 2, .., $l = 1, 2, 3, ...$ When $A = 0$ $X_{nl}$ are the values of the roots of the Bessel functions, and for $B = 0$ $X_{nl}$ are the values of the roots of Neumann functions.

From (14) we obtain the allowed discrete values of the rest masses of the particles

$$m_{nl} = \pm \frac{1}{a} \frac{\hbar}{c} X_{nl}. \qquad (23)$$

Note that when considering the particles in our opinion, there is no reason to use the condition of $R(\rho)$ limitation at zero. Moreover, it is of interest to explore solutions of equations (18, 19) for use as a second condition $R(\rho)$ function collapse at zero

$$|R(\rho)| \to \infty \text{ under } \rho \to 0. \qquad (24)$$

This not strict condition at a singular point does not define the specific values of the expansion coefficients in (20). It requires only the non-zero contribution of Neumann functions in (20). Conditions (21, 24) satisfy the solutions of equations (18,19) in the form

$$F_{nl}(x) = B_{nl}^{(N)} N_n(x) + B_{nl}^{(B)} J_n(x). \qquad (25)$$

In this case, to each pair of $n, l$ numbers can correspond "own" the values of the expansion coefficients, and therefore "own" $F_{nl}(x)$ function. Now in the formula (23) $X_{nl}$ are values of the $F_{nl}(x)$ function roots in (25). That is, the conditions (21, 24) allow a "non-linearity" of additional conditions to them, defining the specific values of the expansion coefficients in (25). These additional conditions may themselves depend on the expansion coefficients in (25), that is on the kind of particle wave function in the state with $n, l$ quantum numbers. Preliminary estimates indicate that perspective for the study is the "non-linear" condition

$$\frac{\partial F_{nl}}{\partial x}\bigg|_{x \to X_{nl}} = (-1)^{l+1} \cdot J_n\left(X_{n1}^{(N)}\right). \qquad (26)$$

Here, $X_{01}^{(N)}$ is the first root of the n-order Neumann function.

It is of interest to attempt to detect peculiarities in the mass spectrum of real elementary particles according to the Helmholtz equation (11), in the simplest case - in accordance with the formulas (23.25). Here are the results of conformity assessment experimental electron, proton, neutron, muon and τ-lepton rest masses to the calculated values using the formula (23) for the same $a$ magnitude. Values were calculated

$$\Delta^{(B)}_{n,l} = |m - m^{(B)}_{nl}|/m, \quad \Delta^{(N)}_{n,l} = |m - m^{(N)}_{nl}|/m, \quad (27)$$

where $m$ is the experimental value of the rest mass of the particles, $m^{(B)}_{nl}$ - the values calculated by the formula (23) for $B^{(N)}_{nl} = 0$, $m^{(N)}_{nl}$ - the values calculated by the formula (23) for $B^{(B)}_{nl} = 0$. That is, an attempt was made to establish compliance with the rest mass of each of the five particles of a certain root of the Bessel and Neumann functions.

When selecting the $a$ value in the form of

$$a = \frac{\hbar}{c} \frac{X^{(N)}_{0,1}}{m_e} \quad (28)$$

(here $m_e$, $m_p$, $m_n$, $m_\mu$, $m_\tau$ are the experimental values of the rest mass of the corresponding particles [10]) we have the following correspondence.

For an electron $\Delta^{(N)}_{0,1} = 0$. For a proton at n = 0, $l$ = 523, $m^{(N)}_{0,523} \approx 938.245\ MeV/c^2$, $\Delta^{(N)}_{0,523} = 2.9 \cdot 10^{-5}$. For a neutron at n = 0, $l$ = 524, $m^{(N)}_{0,524} \approx 940.042\ MeV/c^2$, $\Delta^{(N)}_{0,524} = 5.1 \cdot 10^{-4}$. For a muon at n = 0, $l$ = 59 $m^{(N)}_{0,59} \approx 105.547\ MeV/c^2$, $\Delta^{(N)}_{0,59} = 1.1 \cdot 10^{-3}$. For a τ-lepton at n=0, $l$ = 989, $m^{(B)}_{0,989} \approx 1776.33\ MeV/c^2$, $\Delta^{(B)}_{0,989} = 3.8 \cdot 10^{-4}$.

A comparison of these values, we see that $\Delta^{(N)}_{n,l}$ increase in the set of $e$ - $p$ - $n$ - $\mu$, and the value of the τ-lepton rest mass well corresponds to the root of the Bessel function (the $B^{(N)}_{0,989}$ coefficient in (25) is small). This result, apparently, does not seem to be accidental, and may be associated with the following circumstance. The lifetime of the particles in a row $e$ - $p$ - $n$ - $\mu$ - $\tau$ decreases (an electron is stable, a proton is almost stable, while the lifetimes of the neutron, muon and τ-lepton are 889(3) s; 2,19703(4)·$10^{-6}$ s; 295(3)·$10^{-15}$ s, respectively).

The natural assumption is the following. Stability (lifetime) of the particle depends on the ratio $|B^{(B)}_{nl}/B^{(N)}_{nl}|$ of the expansion coefficients in (25). Since the $\Delta^{(N)}_{nl}$ increase in the set of $e$ - $p$ - $n$ - $\mu$, the difference between the roots of the equation (25), exactly corresponding to the experimental values of the rest masses and the roots of "clean" Neumann functions, and consequently $|B^{(B)}_{nl}/B^{(N)}_{nl}|$ in this series are also increasing. At the same time, for a τ-lepton, which has a small lifetime, $|B^{(N)}_{0,989}/B^{(B)}_{0,989}|$ value there is a small.

## Conclusion

Considered particle described by the wave functions (9) with a dispersion relation (2, 12) is, in general, thought up. Currently, there is no rigorous proof the need to use just such a particular solutions of the relativistic wave equation. Let's note, however, the main "advantages" of such particles. Particles have discrete values of the squares of the rest masses, corresponding to the eigenvalues of the $\hat{p}^2/c^2$ operator. Detection of compliance solutions of (11) in various models to real objects, such as elementary particles or nuclei could be seen as a justification of the form of the wave function (9) for these objects (discussed in this paper is the simplest model). These results indicate such compliance is possible in principle not only for the mass spectrum of elementary particles, but also for their lifetimes, according to the discrete solutions of the Helmholtz equation.